\documentclass[aps,%prl,
floatfix,twocolumn,a4paper]{revtex4}

\usepackage{graphicx,epsf}%
\newcommand {\apgt} {\ {\raise-.5ex\hbox{$\buildrel>\over\sim$}}\ }
\newcommand {\aplt} {\ {\raise-.5ex\hbox{$\buildrel<\over\sim$}}\ }
\usepackage{graphicx}
\usepackage{amssymb}
\usepackage{caption}

\begin{document}

\title{50 years of spin glass theory}

\author{David Sherrington}
\affiliation{Rudolf Peierls Centre for Theoretical Physics, University of Oxford, UK}
\email{david.sherrington@physics.ox.ac.uk}
\author{Scott Kirkpatrick}
\affiliation{School of Engineering and Computer Science, Hebrew University, Jerusalem, Israel}
\email{kirk@cs.huji.ac.il}

%\begin{figure}[t]
%\includegraphics[width=3in]{Figures/SG2everything}
%\caption{
%SK post Parisi;
%$\overline{J_{ij}} =J_{0}/N,  \overline{(J_{ij}^2-{\overline{J_{ij}}^2}}={{J_0}^2}/N   $   }
%\end{figure}

\begin{abstract}

In 1975, two theoretical papers were published that together sparked major new directions, conceptual, mathematical and practically-applicable, in several 
 previously disparate fields of science. In this short comment/review, we expose key aspects of the thinking behind those papers, their implementations and their implications, along with sketches of several subsequent and consequential developments.
 
\end{abstract}

\maketitle

These papers were ``Theory of spin glasses” by Sam Edwards and Phil Anderson (EA)\cite{EA}  and 
``Solvable Model of a Spin-Glass” by David Sherrington and Scott Kirkpatrick (SK)\cite{SK}, both concerned with trying to understand recent experiments that suggested a new phase of matter. 

The experiments, on substitutional alloys of magnetic transition and non-magnetic noble metals, indicated that at low but finite concentration of the magnetic ions and low temperature their magnetic moments (spins) are individually `frozen' in their orientation over nuclear spin resonance timescales, but in a quasi-random, non-periodic, fashion. This state was named {\bf{spin glass}} by analogy with positional amorphousness in normal glasses. Initially, transition to this behaviour as temperature was reduced seemed gradual, but in 1972 Vincent Cannella and John Mydosh \cite{CM} observed that it sharpened as the external field was reduced, suggesting a true cooperative phase transition to a hitherto-unknown phase.

 However, there was no potentially satisfying theoretical explanation until Edwards and Anderson took an interest.  The experimental systems involved quenched-random site-occupation disorder together with long-range separation-decaying and sign-oscillating RKKY spin-pair interactions. EA replaced this by a technically more convenient but conceptually similar model having spins on every site ($i=1...N$) with Hamiltonian  $H=-\sum_{i<j}{J_{ij}  {\bf{S}}_{i}.{\bf{S}}_{j}}$, where $J_{ij}$ is the coupling between nearest-neighbour spins ${\bf{S}}_i$ and ${\bf{S}}_j$, chosen randomly from a Gaussian probability distribution of zero mean and variance $J^2$. They then studied its statistical physics with further inspirational concepts and mathematical tools.
 
Statistical physics interest is in averages, over instances of the quenched parameters \{$J_{ij}$\}, of physical  observables such as the free energy $F=-T\ln{Z}$, where $Z$ is the partition function, which is generally difficult, rather than $Z$ itself, which would be easier but unphysical. To deal with this,
Edwards and Anderson employed a novel but irregular mathematical procedure to effect a transformation of $\ln{Z}$ into a product of $Z$s, using the formal identity 
${\ln{Z} =\lim_{n \rightarrow 0} {n^{-1}\{{Z^n} -1}\}; Z^{n}={\prod_{\alpha=1,..n} }Z_{\alpha}}$ where 
the \{$\alpha$\} label  
`replicas'  (systems with the same quenched Hamiltonian but with their spin variables evolving independently 
of one another).  

Another of their important innovations was the introduction of a new type of order parameter,
 $q =N^{-1} \sum_{i} {\langle {S_{i}^{\alpha} S_{i}^{\beta}} \rangle};{\alpha\neq\beta}$, where $N$ is the number of spins in the system.
With some further assumptions, approximations and ans\"atze,
they were able to demonstrate a phase transition to a new `amorphous' cooperative spin ordering
in which the order parameter $q$ is identifiable as $\overline{  {\langle  {S_{i}} \rangle} ^2   }$.

Their paper immediately excited one of us (D.S.) by both its conceptually new ideas and its new methods of analysis, but he wanted to evaluate the correctness of its assumptions and approximations. To test these further, he devised a related model that he expected should be solvable exactly. The model had an infinite-ranged and range-independent interaction distribution and  also allowed for a non-zero mean for the interaction distribution, in order to  emulate the experimental observation of transitions from ferromagnetism to spin glass as the concentration of magnetic constituent is reduced.
 For calculational simplicity while maintaining conceptual clarity,
 he considered discrete Ising spins, rather than the classical fixed-length vectors of EA. 
 Using the same two ans\"atze as employed by Edwards and Anderson, 
 he derived the Ising analogues of their equations  and, shortly thereafter,  interested the other of us (Scott K) in further joint studies, both analytic and by computer simulation. 

These joint studies yielded a phase diagram in qualitative accord with experiment, but also led to discoveries that had major further consequence; (i) a prediction of a negative entropy at zero temperature, a result that is fundamentally forbidden for discrete variables and signalled a serious procedural error, (ii) evidence of a complex `energy landscape’ with many hills and valleys on many levels, hindering equilibration. 

More  detailed computer simulations, reported in a subsequent longer paper  (KS)\cite{KS} in 1978, indicated a lower ground state energy than the calculation. There was another  problem!

EA and SK had both employed a `natural' ansatz for  $q^{\alpha\beta} =N^{-1} \sum_{i} {\langle {S_{i}^{\alpha} S_{i}^{\beta}} \rangle}$, that it should have the same value $q$ for any non-identical pair of replicas
${\alpha}\neq {\beta}$, a property now known as `replica symmetry'. However, also in 1978,
Jairo de Almeida and David Thouless  \cite{Almeida1978a}  showed that this choice is unstable against fluctuations in replica space. Two further years later, Giorgio Parisi \cite{Parisi1979} devised a revolutional `replica symmetry breaking' 
ansatz that solved the negative entropy problem and also gave a lower ground state energy result, close to that of the simulation reported in Ref \cite{KS}. It involved a radically new mathematical conceptualization and formulation, in which the order parameter is now a function $q(x);0\leq x \leq 1$. The figure illustrates the eventual resultant phase diagram of the SK model. 

\begin{figure}
\includegraphics[width=3.5 in]{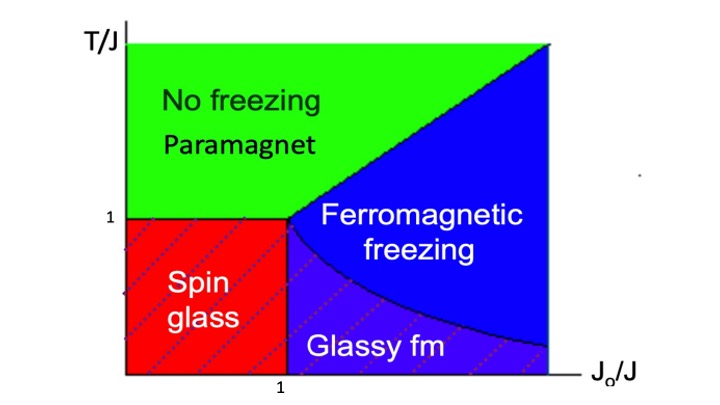}
{\center
\textbf {Full Sherrington--Kirkpatrick model phase diagram}
\newline
Post-Parisi, T=temperature, Hatched = complex/glassy}
\vspace{0.1cm}
$\overline{J_{ij}} =J_{0}/N$,
 $\overline {    J_{ij}^2 - \overline{J_{ij}}^2    }={J^2}/N; \hspace{0.3 cm} N=${\rm number of spins}
\end{figure}

Then, in 1983, Parisi further provided the physical explanation \cite{Parisi1983} of the order parameter, in terms of the co-existence of a great multiplicity of pure equilibrium states, and, in doing so, opened up new insightful physical conceptualization. Interest and activity exploded, leading to many further new discoveries \cite{Mezard_etc}, such as ultrametricity, non-self-averaging, chaotic evolution of the 'free energy landscape', unusual dynamics, breakdown of the usual fluctuation--dissipation theorem, extensions to new theoretical models and further new ideas and perspectives, not only in physics but in many other areas of complexity. Of particular note in the extensions is the discovery of new features in systems lacking the `symmetry of definiteness' between positive and negative interactions of EA and SK, as in Potts  models with Potts-dimension greater than 2 (and especially greater than 4) and systems with more than binary interactions ($p$-spin with $p>2$. These include having (i) different transition temperatures for Gibbs thermodynamics and for dynamics/marginal stability and (ii) different types of replica symmetry breaking, including transitions between them (Gardner \cite{Gardner} transition). This further led to significant new concepts and results concerning simple (nonmagnetic) glasses \cite{Wolynes}
 \cite{Simple_Glasses}. Parisi was awarded the Nobel prize in 2021.

In 1982, John Hopfield \cite{Hopfield}  proposed and demonstrated 
by  computer simulation a model system for distributed attractor memory and associative retrieval, in which the stored memories  can be interpreted as an extension of the Sherrington--Kirkpatrick model to include many gauge-transformed ferromagnets. The retrieval of any one of these memories is analogous to finding the ferromagnetic phase in the SK  spin glass, with the contributions of the other memories interfering analogously to the interaction-variance term in SK and EA, leading to a critical retrieval capacity of order the number of nodes. 
Through an extension of SK analysis, Daniel Amit, Hanoch Gutfreund and Haim Sompolinsky \cite{AGS}  demonstrated its ‘basins of attraction’ of stored memories, their competitively constrained extents and passage to ’spin glass complexity’ beyond the critical capacity.

Both biological brains and artificially intelligent neural networks require mechanisms for learning as well as recalling and generalizing. Within an extension of the SK conceptualization this can be translated to gradually modifying the interactions/synapses in a learning stage, to be later recalled or generalised in faster neural dynamics, either in a recurrent version (as one might model the brain) or a directed layered one (as in most AI/`deep-learning' applications). 

The Edwards--Anderson and Sherrington--Kirkpatrick models are examples of optimization in the presence of 
temperature-noise stochasticity. At the time of our collaboration, Scott, at IBM Research, was involved in understanding problems of designing computer systems and their components.  He recognised that extensions of the computer-simulation methods developed for spin glasses would be a natural and powerful generalization of the highly specialised and more limited approaches then used in the placement of components and the routing of signals between them.  The generalization to other optimization problems led to the concept and application of `simulated annealing' \cite{KGV82}. 

Edwards and Anderson's theoretical replication concept has also led to informative computer-simulation analogues, in which replica systems with identical quenched disorder and the same temperature level of noise are allowed to evolve independently and correlations between them studied to indicate spin glass-like transitions, critical properties and physical measures \cite{Chap5}.

Another consequence within computer science has been the recognition of interest in `typical' as well as `worst-case' studies.

As well as both models stimulating much further research in theoretical condensed matter physics,
the Sherrington--Kirkpatrick model also provided new exciting challenges for mathematical physicists 
\cite{Guerra} and probabilists \cite{Talagrand} \cite{Panchenko} and led to several new developments in both subjects. Its rigorous solution by Michel Talagrand played an important role in his earning the 2024 Abel Prize.  

So, the Sherrington--Kirkpatrick model is indeed `solvable', but the solution has proven to be very subtle, as well as consequential.

Prior to the publication of Ref. {\cite{SK}}, mean field (or infinite-range) theory was mostly considered to be fairly trivial. However, the study of the SK model has demonstrated that it can be complex and highly non-trivial when it involves quenched disorder and frustration (competing interactions or instructions). It has spawned a plethora of conceptual extensions, some physics-orientated but also many in many other areas. 

Returning to materials, the actual alloys that started the journey have not proven practically valuable, but one of us (DS) has argued that the pictures developed provide possible understanding of some other solid state alloys, experimentally discovered long ago and of practical value but with less theoretical investigation, such as relaxor ferroelectrics and martensitic shape-memory alloys.

In conclusion, we  have tried to show briefly how blue sky experimental studies of some alloys at low temperature have, by a combination of theoretical minimalist modelling, unconventional analysis, recognition of puzzling anomalies, curiosity, tenacity and brilliance, led to unanticipated major conceptual/theoretical, mathematical, practically-applicable and computational advances, illustrating the under-appreciated power of theoretical physics for conceptual and mathematical transfers of understanding and applications between often-physically-quite-different areas of science and its application.

\bibliography{Sherrington-Kirkpatrick-references-4}

\end{document}